\documentclass[fleqn,twoside]{article}
\usepackage{espcrc2}
\usepackage{graphicx}

\newcommand{\AmS}{{\protect\the\textfont2
  A\kern-.1667em\lower.5ex\hbox{M}\kern-.125emS}}
\def\myfigure#1{\centerline{\framebox[5cm][c]{FIGURE}}}

\def\refjl#1#2#3#4#5#6{\bibitem{#1} #2, {#3} {#4} (#5) #6.}

\def\etal{{et al}}
\def\NP{Nucl. Phys.}
\def\NPPS{Nucl. Phys. B (Proc. Suppl.)}
\def\PL{Phys. Lett.}
\def\PRL{Phys. Rev. Lett.}
\def\PR{Phys. Rev.}

\def\ZP{Z. Phys.}

\def\RPP{Rep. Prog. Phys.}

\def\PPNP{Prog. Part. Nucl. Phys.}

\newcommand{\eqn}[1]{(\ref{#1})}
\newcommand{\be}{\begin{equation}}
\newcommand{\ee}{\end{equation}}
\newcommand{\no}{\nonumber}
\newcommand{\bel}[1]{\be\label{#1}}
\newcommand{\ba}{\begin{array}{c}}
\newcommand{\bat}{\begin{array}{cc}}
\newcommand{\ea}{\end{array}}
\newcommand{\beqn}{\begin{eqnarray}}
\newcommand{\eeqn}{\end{eqnarray}}

\newcommand{\bi}{\begin{itemize}}
\newcommand{\ei}{\end{itemize}}

\newcommand{\gsim}{~{}_{\textstyle\sim}^{\textstyle >}~}
\newcommand{\lsim}{~{}_{\textstyle\sim}^{\textstyle <}~}

\newcommand{\cL}{{\cal L}}

\title{Leptonic Probes of the Standard Model}

\author{A. Pich\address{
         Departament de F\'{\i}sica Te\`orica,
         IFIC, Universitat de Val\`encia --- CSIC, \\
         Apt. Correus 22085, E--46071 Val\`encia, Spain}}

\begin{document}

\begin{abstract}
\vspace{1pc}
Precise measurements of the lepton properties provide
stringent tests of the Standard Model structure and accurate 
determinations of its parameters.
We overview the present status of a few selected topics: 
lepton universality, QCD
tests and the determination of $\alpha_s$ and $m_s$ from
hadronic $\tau$ decays, the anomalous lepton magnetic moments
and neutrino oscillations.
\end{abstract}

\maketitle

\section{INTRODUCTION}

The Standard Model constitutes one of the most successful 
achievements in modern physics. It is a very elegant
theoretical framework, which is able to describe nearly all
known experimental facts in particle physics
\cite{eso02,GR:02}.

The known leptons have provided clean probes to perform
very precise tests of the electroweak gauge structure,
at the 0.1\% to 1\% level.
Moreover, the hadronic $\tau$ decays turn out to be
a beautiful laboratory for studying strong interaction effects
at low energies \cite{taurev98,taurev00,tau00,Stahl}.
Accurate determinations of the QCD coupling and the strange quark mass
have been obtained with $\tau$ decay data.

Very recently, the first hints of new physics beyond the  
Standard Model have also emerged from the lepton sector.
Convincing evidence of neutrino oscillations has been
obtained by SNO \cite{SNO} and Super-Kamiokande \cite{SKsolar,SKatm}. 
Combined with data from other neutrino experiments, it shows that
$\nu_e\to\nu_{\mu,\tau}$ and $\nu_\mu\to\nu_\tau$
lepton-flavour-violating transitions do occur \cite{concha}.

\section{LEPTON UNIVERSALITY}
\label{sec:universality}

\begin{table}[bth]
\centering
\caption{Present constraints on $|g_l/g_{l'}|$.}
\label{tab:ccuniv}
\vspace{0.2cm}
\begin{tabular}{lc}
\hline
& $|g_\mu/g_e|$ \\ \hline
$B_{\tau\to\mu}/B_{\tau\to e}$ & $0.9999\pm 0.0020$ \\
$B_{\pi\to e}/B_{\pi\to\mu}$ & $1.0017\pm 0.0015$ \\
$B_{W\to\mu}/B_{W\to e}$  & $1.000\pm 0.011$ \\
\hline\hline
& $|g_\tau/g_\mu|$  \\ \hline
$B_{\tau\to e}\,\tau_\mu/\tau_\tau$ & $1.0004\pm 0.0023$ \\
$\Gamma_{\tau\to\pi}/\Gamma_{\pi\to\mu}$ &  $0.9999\pm 0.0036$ \\
$\Gamma_{\tau\to K}/\Gamma_{K\to\mu}$ & $0.979\pm 0.017$ \\
$B_{W\to\tau}/B_{W\to\mu}$  & $1.026\pm 0.014$
\\ \hline\hline
& $|g_\tau/g_e|$  \\ \hline
$B_{\tau\to\mu}\,\tau_\mu/\tau_\tau$ & $1.0002\pm 0.0022$ \\
$B_{W\to\tau}/B_{W\to e}$  & $1.026\pm 0.014$
\\ \hline
\end{tabular}
\end{table}
%

In the Standard Model all lepton doublets have identical couplings to the $W$ boson:
\begin{equation}\label{eq:W_ln}
{\cal L}\; =\; {g\over 2\sqrt{2}}\; W_\mu^\dagger\;\sum_l \; 
\bar\nu_l\,\gamma^\mu (1-\gamma_5)\, l \; +\; \mathrm{h.c.}
\; .
\end{equation}
Comparing the measured decay widths of leptonic or semileptonic decays
which only differ by the lepton flavour, one can test experimentally
that the $W$ interaction is indeed the same, i.e. that \
$g_e = g_\mu = g_\tau \equiv g\, $.
As shown in Table~\ref{tab:ccuniv}, the present data
\cite{PDG,LEPEWWG,Matorras}
verify the universality of the leptonic charged-current couplings
to the 0.2\% level.

The values of $B_{\tau\to\mu/ e}$ and $\tau_\tau$
are already known with a precision of $0.29\%$ and $0.34\%$,
respectively \cite{Matorras}. It remains to be seen whether
BABAR and BELLE could make further improvements.
The $\mu$ lifetime has been measured to a much better precision of
$10^{-5}$. The universality tests require also
a good determination of $m_\tau^5$, which is only known to the
$0.08\%$ level. An improved measurement of the $\tau$ mass could be 
expected from CLEO-C \cite{CLEO-C} and CMD-2 \cite{CMD-2}, 
through a detailed analysis
of $\sigma(e^+e^-\to\tau^+\tau^-)$ at threshold \cite{Pedro}.

\begin{figure}[bt] 
\begin{center}
\includegraphics[width=7.cm]{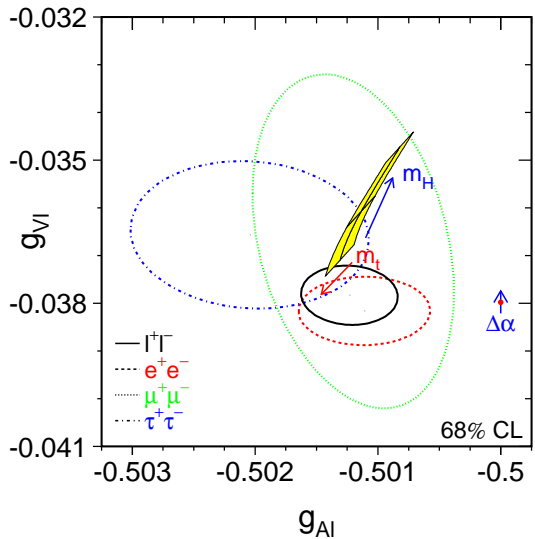}
\end{center}\vskip -1cm
\caption{68\% probability contours in the $a_l$-$v_l$ plane from
LEP and SLD data \cite{LEPEWWG}. 
The shaded region is the Standard Model
prediction
for $m_t =174.3\pm 5.1$~GeV and $m_H = 300\,{}^{+700}_{-186}$~GeV.
}
\label{fig:gal_gvl_contours}
\end{figure}

The interactions of the neutral $Z$ boson are diagonal in flavour.
Moreover, all fermions with equal electric charge have identical
vector, $v_f = T_3^f\, (1-4\, |Q_f|\,\sin^2{\theta_W})$, and
axial-vector, $a_f = T_3^f=\pm 1/2$, couplings to the $Z$.
This has been accurately tested at LEP and SLD through a
precise analysis of $e^+e^-\to\gamma,Z\to f\bar f$ data.
Figure~\ref{fig:gal_gvl_contours} shows the 68\% probability 
contours in the
$a_l$-$v_l$ plane, obtained from leptonic observables \cite{LEPEWWG}.
The universality of the leptonic $Z$ couplings is now verified to 
the $0.15\%$
level for $a_l$, while only a few per cent precision has been achieved 
for $v_l$ due to the smallness of the leptonic vector coupling.
Assuming universality, 
the measured leptonic asymmetries provide
an accurate determination of the electroweak mixing angle \cite{LEPEWWG}:
\begin{equation}
\sin^2{\theta_W}\; = \; 0.23113\pm 0.00021\; .
\end{equation}

\section{HADRONIC TAU DECAYS}
\label{sec:hadronic}

The semileptonic decay modes $\tau^-\to\nu_\tau H^-$ probe the  matrix
element of the left--handed charged current between the vacuum and the
final hadronic state $H^-$.

For the decay modes with lowest multiplicity,
$\tau^-\to\nu_\tau\pi^-$, and $\tau^-\to\nu_\tau K^-$, the  relevant
matrix  elements  are  already  known  from  the  measured  decays
$\pi^-\to\mu^-\bar\nu_\mu$  and  $K^-\to\mu^-\bar\nu_\mu$.
The corresponding $\tau$ decay widths can then be accurately predicted.
As shown in Table~\ref{tab:ccuniv}, these predictions are in good
agreement with the measured values, and provide a quite precise test
of charged--current universality.

For the two--pion final state, the hadronic matrix element is
parameterized in terms of the so-called pion form factor \
[$s\equiv (p_{\pi^-}\! + p_{\pi^0})^2$]:
\bel{eq:Had_matrix}
\langle \pi^-\pi^0| \bar d \gamma^\mu  u | 0 \rangle \equiv
\sqrt{2}\, F_\pi(s)\, \left( p_{\pi^-}- p_{\pi^0}\right)^\mu \, .
\ee
A dynamical understanding of the pion form factor can be achieved
\cite{GP:97,DPP:00,Portoles,Juanjo},
by using analyticity, unitarity and some general properties of QCD.

At low momentum transfer, the coupling of any
number of $\pi $'s, $K$'s and $\eta$'s to the
V$-$A current can be rigorously calculated with
Chiral Perturbation Theory \cite{GL:85,EC:95} techniques,
as an expansion in powers of $s$ and light quark masses over the
chiral symmetry breaking scale ($\Lambda_\chi\sim 1$ GeV).
This includes chiral loop corrections, which encode the absorptive
contributions required by unitarity.
The short--distance information is contained in the so-called
chiral couplings, which are known to be dominated by the effect
of the lowest--mass resonances \cite{EGPR:89}.

\begin{figure}[tbh]
\includegraphics[angle=-90,width=7.5cm,clip]{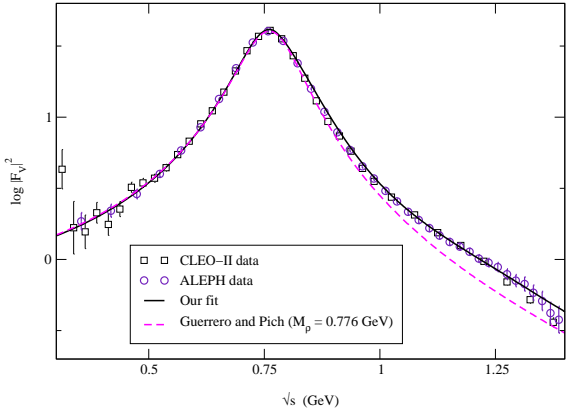}
\vspace{-.5cm}
\caption{Pion form factor data \cite{ALEPHpiff,CLEOpiff}
compared with theoretical
predictions \protect\cite{Portoles}.}
\label{fig:pionth}
\end{figure}

In the limit of an infinite number of quark colours $N_C$, QCD reduces
to a theory of tree-level resonance exchanges \cite{HO:74,Tempe}.
Thus, the $\rho$ propagator governs the pion form factor
at $\sqrt{s}\lsim 1$ GeV, providing an all-order
resummation of the polynomial chiral corrections.
The relevant $\rho$ couplings are determined requiring 
$F_\pi(s)$ to satisfy the correct QCD behaviour at large $s$ \cite{EGPR:89}.
The leading $1/N_C$ corrections correspond to
pion loops and can be incorporated by matching the
large--$N_C$ result with the Chiral Perturbation Theory description
\cite{GP:97}. Using analyticity and unitarity constraints,
the chiral logarithms associated with those pion loops
can be exponentiated to all orders in the chiral expansion. 
Putting all these
fundamental ingredients together, one gets the result \cite{GP:97}:
$$
F_\pi(s) = {M_\rho^2\over M_\rho^2 - s - i M_\rho \Gamma_\rho(s)}
\exp{\left\{-{s \,\mbox{\rm Re} A(s)           
\over 96\pi^2f_\pi^2} \right\}} ,
$$
where
$$
A(s) \equiv \log{\left({m_\pi^2\over M_\rho^2}\right)} +
8 {m_\pi^2 \over s} - {5\over 3} +
\sigma_\pi^3 \log{\left({\sigma_\pi+1\over \sigma_\pi-1}\right)}
$$
contains the one-loop chiral logarithms,
$\sigma_\pi\equiv\sqrt{1-4m_\pi^2/s}$ and the
off-shell $\rho$ width \cite{GP:97,DPP:00} is given by
\be 
\Gamma_\rho(s)\, =\, \theta(s-4m_\pi^2)\,\sigma_\pi^3\,
M_\rho\, s/(96\pi f_\pi^2)\, .
\ee
This prediction, which only depends on $M_\rho$,
$m_\pi$ and the pion decay constant
$f_\pi$, is compared with the data
in Fig.~\ref{fig:pionth}. The agreement is
rather impressive and extends to negative $s$ values, where
the $e^-\pi$ elastic data (not shown in the figure) sits.

The small effect of higher $\rho$ resonance contributions and additional
next-to-leading in $1/N_C$ corrections can be easily included,
at the price of having some free parameters
which decrease the predictive power \cite{Portoles,Juanjo}.
This gives a better description of the $\rho'$ shoulder around 1.2 GeV.

The dynamical structure of other hadronic final states can be
investigated in a similar way.
The more involved $\tau\to\nu_\tau 4\pi$ and $e^+e^-\to 4\pi$
transitions have been already studied recently \cite{EU:02}.
A theoretical analysis of the $\tau\to\nu_\tau 3\pi$
data \cite{CLEO3pi,OPAL3pi} is in progress \cite{DPP:01}.

\section{LEPTON ANOMALOUS MAGNETIC MOMENTS}
\label{sec:g-2}

The most stringent QED test \cite{KI:96,KN:02,CM:01}
comes from the high-precision
measurements of the $e$ \cite{VDSD:87} and $\mu$ \cite{BNL:E821}
anomalous magnetic moments  $a_l\equiv (g^\gamma_l-2)/2$:
\beqn\label{eq:a_e}
a_e &\!\!\! =&\!\!\! (115 \, 965 \, 218.69\pm 0.41) \,\cdot\, 10^{-11} \, ,
\\[5pt] \label{eq:a_mu}
a_\mu &\!\!\! =&\!\!\! (116 \, 592 \, 030\pm 80) \,\cdot\, 10^{-11} \, .
\eeqn

To a measurable level, $a_e$ arises entirely from virtual electrons and 
photons; these contributions are known \cite{KI:96,KN:02} to $O(\alpha^4)$.
The impressive agreement achieved between theory and experiment  
has promoted QED to the level of the best theory ever built 
to describe nature.
The theoretical error is dominated by the uncertainty in the 
input value of the QED coupling $\alpha$.
Turning things around, $a_e$ provides the most precise 
determination of the fine structure constant: 
\be
\alpha^{-1} = 137.035\, 998\, 76 \,\pm\, 0.000\, 000\, 52\, .
\ee
The central value is slightly smaller (by $0.82\cdot 10^{-6}$)
than the usually quoted value \cite{KI:96}, due to a small error
recently discovered in the $O(\alpha^4)$ contribution to $a_l$
\cite{KN:02}.
 
The anomalous magnetic moment of the muon is sensitive to small
corrections from virtual heavier states; compared to $a_e$, they scale 
as $m_\mu^2/m_e^2$. The Standard Model prediction can be decomposed
in five types of contributions:
\beqn
10^{11} \cdot a_\mu^{\mathrm{th}} \, = \,
116\; 584\; 718\;\pm\; \phantom{8}3&& \!\!\!\!
\mathrm{QED}\:\mbox{\cite{KI:96,KN:02,CM:01}}
\no\\ \mbox{} +\phantom{6\;}152\;\pm\; \phantom{8}4 &&
\!\!\!\!\mathrm{EW}\:\cite{CM:01,a_mu_EW}
\no\\ \mbox{} +
6\; 943\;\pm\; 85 &&\!\!\!\!\mathrm{had, LO}\:\cite{DEHZ:02}
\no\\ \mbox{} -\phantom{6\;}100\;\pm\; \phantom{8}6&&
\!\!\!\!\mathrm{had, NLO} \:\cite{KR:97}
\no\\ \mbox{} +\phantom{6\; 0} 90\;\pm\;  40 &&
\!\!\!\!\mathrm{lbl}\:\cite{light_new,light_old}
\no\\  
= \, 116\; 591 \; 803\;\pm\; 94 && \hskip -.8cm .
\no\eeqn
This result differs by $1.8\,\sigma$ from the experimental
value \eqn{eq:a_mu}.

\begin{figure}[tbh]
\includegraphics[width=7.5cm]{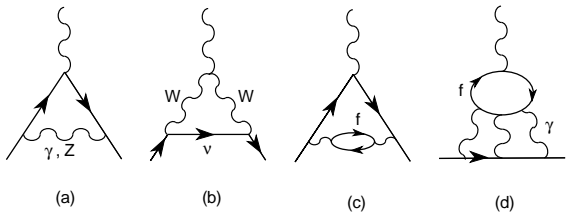}
\vspace{-.75cm}
\caption{Feynman diagrams contributing to $a_l$.}
\label{fig:AnMagMom}
\end{figure}

The main theoretical uncertainty on $a_\mu$ has a QCD origin. 
Since quarks have electric charge, virtual quark-antiquark pairs 
induce
{\it hadronic vacuum polarization} corrections to the photon propagator 
(Fig.~\ref{fig:AnMagMom}.c).
Owing to the non-perturbative character of QCD at low energies, 
the light-quark contribution cannot be reliably calculated at 
present. Fortunately, this effect can be 
extracted from the measurement of the cross-section 
$\sigma(e^+e^-\to \mbox{\rm hadrons})$ 
and from the invariant-mass distribution of the final hadrons in 
$\tau$ decays \cite{Portoles,JE:01}.
The largest contribution comes from the $2\pi$ final state.
The $\tau$ decay determination includes a careful
investigation of isospin breaking effects \cite{CEN:02}, using
the pion form factor expression of ref.~\cite{GP:97}, which
amount to \cite{DEHZ:02} 
$\Delta a_\mu^{\mathrm{had}} = -(112\pm 28)\cdot 10^{-11}$.

At present, there is a discrepancy between the $2\pi$ contributions
extracted from $e^+e^-$ and $\tau$ data, which translates into
slightly different predictions ($1.8\,\sigma$) for $a_\mu$. 
One gets \cite{DEHZ:02}
$a_\mu^{\mathrm{had,LO}} = (6847\pm 70)\cdot 10^{-11}$
from $e^+e^-$, while the $\tau$ data gives
$a_\mu^{\mathrm{had,LO}} = (7019\pm 62)\cdot 10^{-11}$,
in better agreement with the BNL-821 measurement of $a_\mu$ \cite{BNL:E821}.
In order to quote a reference number for $a_\mu^{\mathrm{th}}$,
I have used a weighted average of these two determinations, increasing
the error with the appropriate scale factor \cite{PDG}.
New precise $e^+e^-$ and $\tau$ data sets are needed to settle the
true value of $a_\mu^{\mathrm{had,LO}}$ \cite{DEHZ:02,German}.

Additional QCD uncertainties stem from the smaller 
{\it light-by-light scattering} contributions (Fig.~\ref{fig:AnMagMom}.d).
A recent reevaluation of these corrections \cite{light_new} has detected 
a sign mistake in previous calculations \cite{light_old}, improving the
agreement with the experimental measurement \cite{BNL:E821}.

The Brookhaven $(g-2)$ experiment \cite{BNL:E821} is expected to push
its sensitivity to at least $4\cdot 10^{-10}$,
and thereby observe the contributions from virtual $W^\pm$ and $Z$ bosons 
\cite{CM:01,a_mu_EW}.
A meaningful test of the electroweak Standard Model contributions
would require a better control of the QCD corrections.

\section{THE TAU HADRONIC WIDTH}
\label{sec:hadronic_width}

The inclusive character of the total $\tau$ hadronic width
renders possible an accurate calculation of the ratio
\cite{BR:88,NP:88,BNP:92,LDP:92a,QCD:94}
%
\bel{eq:r_tau_def}
R_\tau \equiv { \Gamma [\tau^- \rightarrow \nu_\tau
                   \,\mbox{\rm hadrons}\, (\gamma)] \over
                         \Gamma [\tau^- \rightarrow
                \nu_\tau e^- {\bar \nu}_e (\gamma)] }\, ,
\ee
using analyticity constraints and the Operator Product Expansion.
One can separately compute the contributions associated with
specific quark currents:
\be\label{eq:r_tau_v,a,s}
 R_\tau \, = \, R_{\tau,V} + R_{\tau,A} + R_{\tau,S}\, .
\ee
$R_{\tau,V}$ and $R_{\tau,A}$ correspond to the Cabibbo--allowed
decays through the vector and axial-vector currents, while
$R_{\tau,S}$ contains the remaining Cabibbo--suppressed contributions.

The theoretical prediction for $R_{\tau,V+A}$ can be expressed as \cite{BNP:92}
$$
R_{\tau,V+A} = N_C\, |V_{ud}|^2\,
S_{\mathrm{EW}} \left\{ 1 + \delta_{\mathrm{EW}}' +\delta_{\mathrm{P}} 
+ \delta_{\mathrm{NP}} \right\} ,
$$
with $N_C=3$.
The factors $S_{\mathrm{EW}}=1.0194$ and $\delta_{\mathrm{EW}}'=0.0010$
contain the electroweak corrections at the leading
\cite{MS:88} and next-to-leading \cite{BL:90} logarithm
approximation.
The dominant correction ($\sim 20\%$) is the purely perturbative
contribution $\delta_{\mathrm{P}}$, which is fully 
known to $O(\alpha_s^3)$ \cite{BNP:92} and includes a resummation
of the most important higher-order corrections \cite{LDP:92a}.
%

%
\begin{figure}[tbh]
\label{fig:alpha_s}
\includegraphics[width=7.5cm]{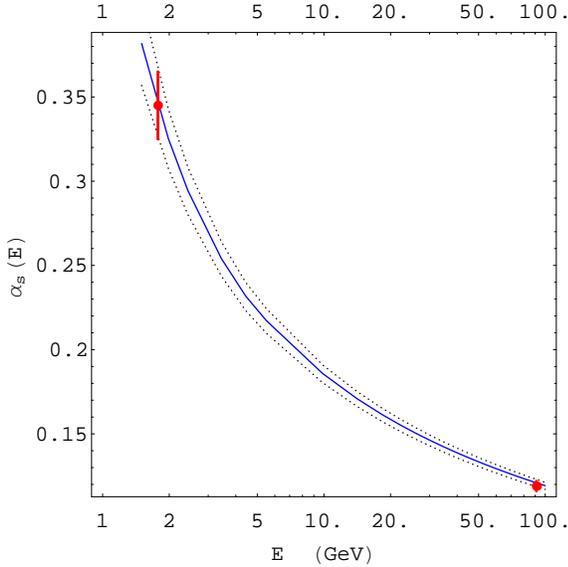}
\vspace{-0.5cm}
\caption{Measured values of $\alpha_s$ in $\tau$ and $Z$ decays.
The curves show the energy dependence predicted by QCD, using
$\alpha_s(m_\tau^2)$ as input.}
\end{figure}

Non-perturbative contributions are
suppressed by six powers of the $\tau$ mass \cite{BNP:92}
and, therefore, are very small. Their numerical size has been
determined from the invariant--mass distribution of the final hadrons
in $\tau$ decay, through the study of weighted integrals \cite{LDP:92b},
\be
R_{\tau}^{kl} \equiv \int_0^{m_\tau^2} ds\,
\left(1 - {s\over m_\tau^2}\right)^k\, \left({s\over m_\tau^2}\right)^l\,
{d R_{\tau}\over ds} \, ,
\ee
which can be calculated theoretically in the same way as $R_{\tau}$.
The predicted suppression \cite{BNP:92}
of the non-perturbative corrections has been confirmed by
ALEPH \cite{ALEPH:98}, CLEO \cite{CLEO:95} and OPAL \cite{OPAL:98}.
The most recent analyses \cite{ALEPH:98,OPAL:98} give
\bel{eq:del_np}
\delta_{\mathrm{NP}} \, =\, -0.003\pm 0.003 \, .
\ee

The QCD prediction
for $R_{\tau,V+A}$ is then completely dominated by the
perturbative contribution; non-perturbative effects being smaller
than the perturbative uncertainties from uncalculated higher-order
corrections. The result turns out to be
very sensitive to the value of $\alpha_s(m_\tau^2)$, allowing for an accurate
determination of the fundamental QCD coupling.
The experimental measurement \cite{ALEPH:98,OPAL:98}
$R_{\tau,V+A}= 3.484\pm0.024$ implies $\delta_{\mathrm{P}} = 0.200\pm 0.013$,
which corresponds (in the $\overline{\rm MS}$ scheme) to
\be\label{eq:alpha}
\alpha_s(m_\tau^2)  =  0.345\pm 0.020 \, .
\ee

The strong coupling measured at the $\tau$ mass
scale is significantly larger than 
the values obtained at higher energies.
From the hadronic decays of the $Z$, one gets
$\alpha_s(M_Z^2) = 0.119\pm 0.003$ \cite{LEPEWWG}, 
which differs from the $\tau$ decay
measurement by eleven standard deviations.
After evolution up to the scale $M_Z$ \cite{Rodrigo:1998zd},
the strong coupling constant in \eqn{eq:alpha} decreases to
\be\label{eq:alpha_z}
\alpha_s(M_Z^2)  =  0.1208\pm 0.0025 \, ,
\ee
in excellent agreement with the direct measurements at the $Z$ peak
and with a similar accuracy.
The comparison of these two determinations of $\alpha_s$ in two extreme
energy regimes, $m_\tau$ and $M_Z$, provides a beautiful test of the
predicted running of the QCD coupling;
i.e. a very significant experimental verification of {\it asymptotic freedom}.

\section{THE STRANGE QUARK MASS}
\label{sec:ms}

The LEP experiments and CLEO have performed an extensive investigation
of kaon production in $\tau$ decays \cite{tau00,PDG}.
ALEPH has determined the inclusive invariant mass
distribution of the final hadrons in the Cabibbo--suppressed decays
\cite{ALEPHms}.
A similar analysis from OPAL should appear soon \cite{OPALms}.
The separate measurement of the $|\Delta S|=0$ and $|\Delta S|=1$
decay widths allows us to pin down the SU(3) breaking effect induced
by the strange quark mass, through the differences:
\beqn
\lefteqn{\delta R_\tau^{kl}  \equiv
  {R_{\tau,V+A}^{kl}\over |V_{ud}|^2} - {R_{\tau,S}^{kl}\over |V_{us}|^2}
  } &&
\\ & &\approx  24\, {m_s^2(m_\tau^2)\over m_\tau^2} \, \Delta_{kl}(\alpha_s)
    - 48\pi^2\, {\delta O_4\over m_\tau^4} \, Q_{kl}(\alpha_s)
\, .\no
\eeqn

The perturbative QCD corrections $\Delta_{kl}(\alpha_s)$ 
and $Q_{kl}(\alpha_s)$ are known \cite{PP:99}
to $O(\alpha_s^3)$ and $O(\alpha_s^2)$,
respectively. The small non-perturbative contribution,
$\delta O_4 \equiv\langle 0| m_s \bar s s - m_d \bar d d |0\rangle
 = -(1.5\pm 0.4)\times 10^{-3}\;\mbox{\rm GeV}^4$,
has been estimated with Chiral Perturbation Theory techniques \cite{PP:99}.
Table~\ref{tab:ms}
shows the measured differences
$\delta R_\tau^{kl}$ \cite{ALEPHms,ChDGHPP:01}
and the corresponding ($\overline{\rm MS}$)
values \cite{ChDGHPP:01} of $m_s(m_\tau^2)$.
The theoretical errors are dominated by the very
large perturbative uncertainties of $\Delta_{kl}(\alpha_s)$
\cite{PP:99,CK:93,CH:97,MA:98,ChKP:98}.

%
\begin{figure}[tbh]\centering
\includegraphics[width=6.5cm]{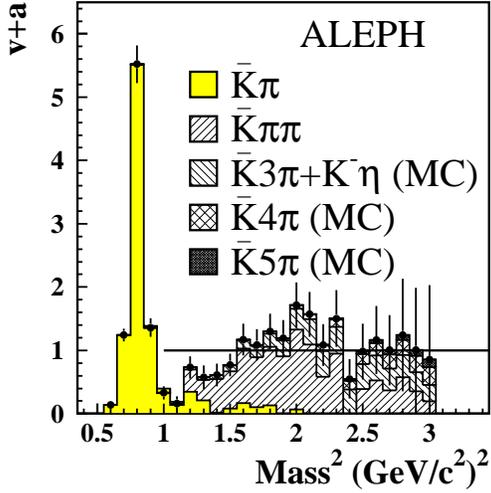}
\vspace{-0.5cm}
\caption{$|\Delta S|=1$ spectral function \protect\cite{ALEPHms}.}
\label{fig:S=1sf}
\end{figure}

\begin{table}[tb]
\caption{Measured moments $\delta R_\tau^{kl}$
\protect\cite{ALEPHms,ChDGHPP:01}  and corresponding $m_s(m_\tau^2)$ values
\protect\cite{ChDGHPP:01}.}
\label{tab:ms}
\begin{tabular}{c|c|c}
$(k,l)$ & $\delta R_\tau^{kl}$ & $m_s(m_\tau^2)$ \ (MeV)
\\ \hline
$(0,0)$ & $0.374\pm 0.133$ & $132\pm 29_{\mathrm{exp}}\pm 14_\mathrm{th}$
\\
$(1,0)$ & $0.398\pm 0.078$ & $120\pm 16_\mathrm{exp}\pm 16_\mathrm{th}$
\\
$(2,0)$ & $0.399\pm 0.054$ & $117\pm 12_\mathrm{exp}\pm 21_\mathrm{th}$
\end{tabular}
\end{table}

A global analysis, using the information from the three moments and
taking into account the strong error correlations,
gives the result \cite{ChDGHPP:01}
$$
m_s(m_\tau^2) = \left(120 \pm 11_{\mathrm{exp}}\pm 8_{V_{us}}
\pm 19_{\mathrm{th}}\right)\;\mbox{\rm MeV} \, .
$$
This corresponds to
$m_s(\mu^2) = (116\,{}^{+20}_{-25})$~MeV at $\mu=2$~GeV.
A similar result is obtained from an analysis based on ``optimal moments''
with improved perturbative convergence \cite{KM:00}.

%
\begin{figure}[tbh]\centering
\includegraphics[width=7.3cm]{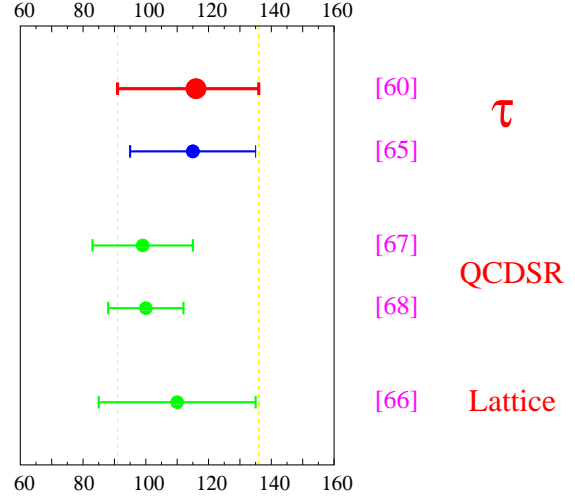}
\vspace{-0.5cm}
\caption{$m_s(\mu^2)$ determinations at $\mu = 2$~GeV.}
\label{fig:ms}
\end{figure}

Figure \ref{fig:ms} compares the $m_s$ value obtained from 
$\tau$ decays with other recent
estimates, at the reference scale $\mu=2$~GeV.
There is a rather large spread of lattice results, 
but the quoted average \cite{LU:01} agrees with the $\tau$
determination.
The latest QCD sum rules results \cite{JOP:02,KM:01} are also
in nice agreement with the $\tau$ value. 
The advantage of the $\tau$ decay width
is the direct use of experimental input, which makes easier to quantify
the associated uncertainties. It suffers, however, from a large theoretical
uncertainty associated with the bad perturbative behaviour of the
scalar contributions to $\Delta_{kl}(\alpha_s)$
\cite{PP:99,CK:93,CH:97,MA:98,ChKP:98}.
This could be improved subtracting theoretically the $J=0$ piece,
with the help of $K\pi$ scattering data \cite{JOP:00}.

The strange quark mass is an important
input to predict the 
CP-violating ratio $\varepsilon'_K/\varepsilon^{\phantom{'}}_K$
of $K\to 2\pi$ decays.
Therefore, it would be very useful to achieve an
improved determination of $m_s$ with the BABAR and BELLE $\tau$
data samples.
Taking
$(m_s + m_{u,d})(\mu^2) = (113\pm 18)$~MeV, at $\mu = 2$~GeV, 
the Standard Model prediction of $\varepsilon'_K/\varepsilon^{\phantom{'}}_K$
is found to be \cite{SPP:01}
\be
\mathrm{Re} (\varepsilon'_K/\varepsilon^{\phantom{'}}_K) =
(1.7\pm 0.2\, {}^{+0.8}_{-0.5}\pm 0.5)\cdot 10^{-3}\, ,
\ee
where the largest error originates in the present uncertainty on $m_s$.
This prediction agrees very well with the measured value
\cite{NA48,NA31,KTeV,E731}
$\mathrm{Re} (\varepsilon'_K/\varepsilon^{\phantom{'}}_K) =
(1.66\pm 0.16)\cdot 10^{-3}$.

\section{NEUTRINO OSCILLATIONS}
\label{sec:oscillations}

%
\begin{figure}[tbh]
\includegraphics[width=7.5cm]{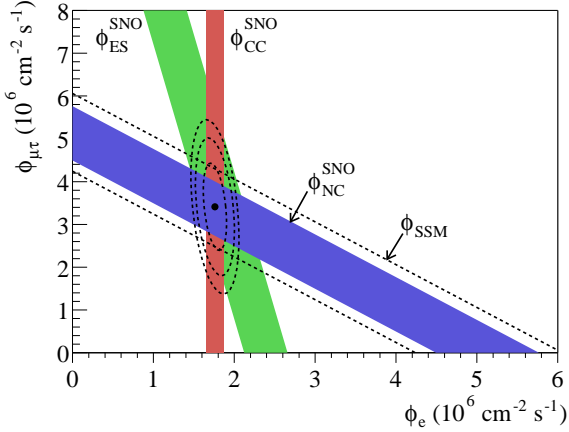}
\vspace{-0.5cm}
\caption{Measured flux of ${}^8B$ solar neutrinos of $\nu_\mu$ or
$\nu_\tau$ type ($\phi_{\mu,\tau}$) versus the flux of $\nu_e$
($\phi_e$)~\cite{SNO}.}
\label{fig:SNO}
\end{figure}

The so-called ``solar neutrino problem'' has been a long-standing
question, since the very first chlorine experiment at the Homestake mine
\cite{DA:68}. The flux of solar $\nu_e$ neutrinos reaching the earth
has been measured by several experiments \cite{concha}
to be significantly below the standard solar model prediction \cite{BA:02}.
The large Super-Kamiokande (SK) experiment \cite{SKsolar}, with its real time
detection technique and direction sensitivity, has fully confirmed the
deficit of solar electron neutrinos.

Very recently, the Sudbury Neutrino Observatory (SNO) has provided
strong evidence that neutrinos do change flavour as they propagate
from the core of the Sun \cite{SNO}, independently of solar model
flux predictions. SNO is able to detect neutrinos through three
different reactions:
\beqn\label{eq:SNOreac}
\nu_e \; + \; d \;\to\; p\; +\; p\; +\; e^- && \mathrm{(CC)}\, ,
\no\\
\nu_x \; + \; d \;\to\; p\; +\; n\; +\; \nu_x && \mathrm{(NC)}\, ,
\\
\nu_x \; + \; e^- \;\to\;  \nu_x \; + \; e^-\hskip .4cm
 && \mathrm{(ES)}\, .\no
\eeqn
While the charged current (CC) reaction is only sensitive to $\nu_e$, 
the neutral current (NC) process
has equal probability for all active neutrino flavours. 
The elastic scattering (ES)
is also sensitive to $\nu_\mu$ and $\nu_\tau$, although the corresponding
cross section is a factor $6.48$ smaller than the $\nu_e$ one.
The measured neutrino fluxes, shown in Fig.~\ref{fig:SNO}, demonstrate
the existence of a non-$\nu_e$ component in the solar neutrino flux
at the $5.3\,\sigma$ level:
\bel{eq:SNOflux}
\phi_{\mu\tau} = \left( 3.41\pm 0.45\, {}^{+0.48}_{-0.45}\right)\cdot
10^6\:\mathrm{cm}^{-2} \mathrm{s}^{-1}\, .
\ee
The total neutrino flux measured with the NC process is consistent
with the solar model prediction \cite{BA:02} 
(dotted line in Fig.~\ref{fig:SNO}).

%
\begin{figure}[tbh]\centering
\includegraphics[width=7.5cm]{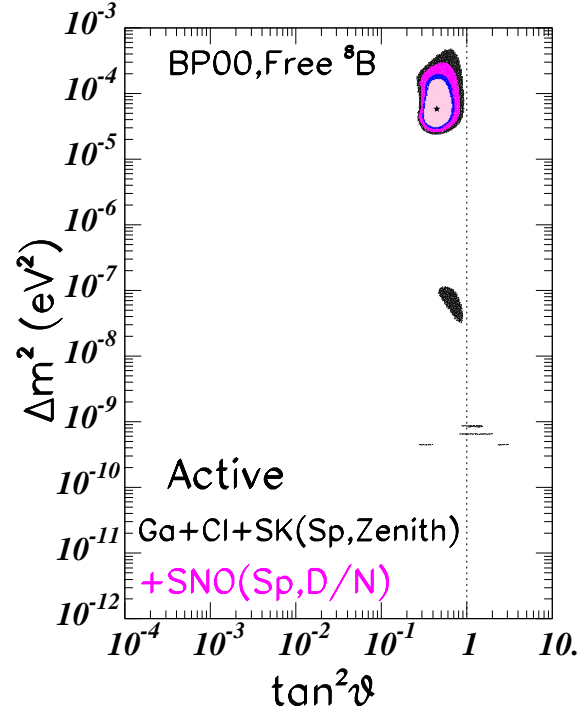}
\vspace{-0.5cm}
\caption{Allowed oscillation parameters (at 90, 95, 99 and 99.7\% CL)
from solar neutrinos \cite{concha,BGP:02}.
The best fit point is marked with a star.}
\label{fig:SolarNu}
\end{figure}

%
\begin{figure}[tbh]\centering
\includegraphics[width=7.5cm,clip]{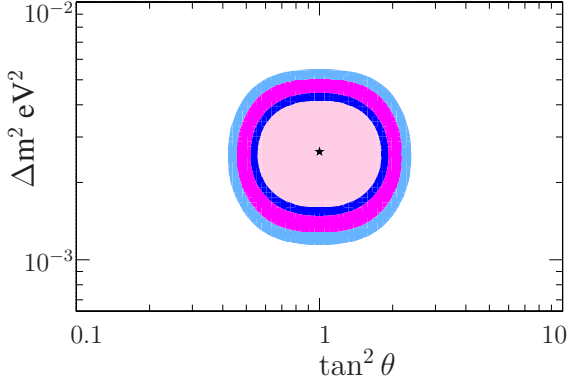}
\vspace{-0.5cm}
\caption{Allowed regions (at 90, 95, 99 and 99.7\% CL) for
$\nu_\mu\to\nu_\tau$ oscillations, from atmospheric neutrino data
\cite{concha}.}
\label{fig:AtmosNu}
\end{figure}

Another very important evidence of neutrino oscillations has been
provided by the SK measurements of atmospheric neutrinos \cite{SKatm}.
The known discrepancy between the experimental observations and the
predicted ratio of muon to electron neutrinos has become much stronger
with the high precision and large statistics of SK.
The atmospheric anomaly has been identified to originate in a reduction
of the $\nu_\mu$ flux, and the data strongly favours the
$\nu_\mu\to\nu_\tau$ hypothesis.
This result should be confirmed soon by K2K \cite{K2K} and further
studied at MINOS \cite{MINOS}.
The direct detection of the produced $\nu_\tau$ is the main goal
of the CERN to Gran Sasso neutrino program \cite{CGSNP}.

Figures \ref{fig:SolarNu} and \ref{fig:AtmosNu} show the present
information on neutrino oscillations, from solar and atmospheric
experiments.
A global analysis, combining the full set of solar, atmospheric and reactor
neutrino data, leads to the following preferred ranges for the
oscillation parameters \cite{concha}:
\beqn\label{nu_mix}
2.4 \cdot 10^{-5}\; <\;\Delta m^2_{21}\, /\, \mathrm{eV}^2
\; <\; 2.4 \cdot 10^{-4}\, ,
&&\no\\
1.4 \cdot 10^{-3}\; <\;\Delta m^2_{32}\, /\, \mathrm{eV}^2
\; <\; 6.0 \cdot 10^{-3} \, ,
&&\no\\
0.27 \; <\;\tan^2{\theta_{12}}\; <\; 0.77 \, ,\hskip 2.2cm
&&\\
0.4 \; <\;\tan^2{\theta_{23}}\; <\; 3.0 \, ,\hskip 2.5cm
&&\no\\
\sin^2{\theta_{13}}\; <\; 0.06 \, ,\hskip 3.5cm
&&\no
\eeqn
where $\Delta m^2_{ij}\equiv m^2_i - m^2_j$ are the mass squared
differences between the neutrino mass eigenstates $\nu_{i,j}$
and $\theta_{ij}$ the corresponding mixing angles (in the standard
three-flavour parameterization \cite{PDG}).
The angle $\theta_{13}$ is strongly constrained by the CHOOZ reactor
experiment \cite{CHOOZ}. In the limit $\theta_{13}=0$, solar and
atmospheric neutrino oscillations decouple because
$\Delta m^2_\odot \ll\Delta m^2_\mathrm{atm}$. Thus, 
$\Delta m^2_{21}$, $\theta_{12}$ and $\theta_{13}$ are constrained by
solar data, while atmospheric experiments constrain
$\Delta m^2_{32}$, $\theta_{23}$ and $\theta_{13}$.

\section{NEW PHYSICS}

The non-zero value of neutrino masses constitutes a clear indication
of new physics beyond the Standard Model framework.
The simplest modification would be to add the needed right-handed
neutrino components to allow for Dirac neutrino mass terms,
generated through the electroweak spontaneous symmetry breaking
mechanism. However, those $\nu_{iR}$ fields would be
$SU(3)_C\otimes SU(2)_L\otimes U(1)_Y$ singlets and, therefore, would
not have any Standard Model interaction (sterile neutrinos).
If such objects do exist, it would seem natural to expect that they
are able to communicate with the rest of the world through some still
unknown dynamics. Moreover, the Standard Model gauge symmetry
would allow for a right-handed Majorana neutrino mass term of arbitrary
size, not related to the ordinary Higgs mechanism. Clearly, new physics
is called for. If the Majorana masses are well above the electroweak
symmetry breaking scale, the see-saw mechanism \cite{RGMY:79}
leads to three light neutrinos at low energies.

We can make a more general analysis without any assumption about
the existence of right-handed neutrinos or any other new particles
at higher scales.
Adopting an effective field theory language, one can
write the most general $SU(3)_C\otimes SU(2)_L\otimes U(1)_Y$
invariant lagrangian, in terms of the known low-energy fields
(left-handed neutrinos only):
\bel{eq:EffL}
\cL\; =\; \cL_\mathrm{SM}\; +\; \sum_{d>4}\; 
{c_d\over\Lambda^{d-4}}\; O_d \, .
\ee
The lagrangian is ordered in terms of increasing dimensionality $d$;
higher-dimension operators being suppressed by higher inverse powers
of the new-physics scale $\Lambda$.
The Standard Model is the unique answer with $d=4$. The first
contributions from new physics appear at $d=5$  and have also a unique
form \cite{WE:79}:
\bel{eq:WE}
\Delta \cL\; =\; - {c_{ij}\over\Lambda}\; \bar L_i\,\tilde\phi\,
\tilde\phi^t\, L_j^c \; + \; \mathrm{h.c.}\, ,
\ee
where $\phi$ and $L_i$ are the scalar and $i$-flavoured lepton 
$SU(2)_L$ doublets,
$\tilde\phi \equiv i\,\tau_2\,\phi^*$ and
$L_i^c \equiv \mathcal{C} \bar L_i^t$.
A similar operator with quark fields is forbidden by the gauge
symmetry, due to the different hypercharges.
After spontaneous symmetry breaking,
$<\phi^{(0)}> = v/\sqrt{2}$, the $d=5$ operator generates a
Majorana mass term:
\bel{eq:Majorana}
\cL_M = -{1\over 2}\, \bar\nu_{iL} M_{ij}\, \nu_{jL}^c
 +\mathrm{h.c.} ,
\quad
M_{ij} = {c_{ij}\, v^2\over\Lambda} .
\ee

Thus, Majorana neutrino masses should be expected on general
symmetry grounds. The relation~(\ref{eq:Majorana}) is
nothing else that the see-saw mechanism
($m_{\nu_{L}}\sim m^2/\Lambda$).
Taking $m_\nu\gsim 0.05$~eV, as suggested by atmospheric neutrino data,
one gets $\Lambda/c_{ij}\lsim 10^{15}$~GeV, amazingly close to the
expected scale of Gran Unification.

With non-zero neutrino masses,
the leptonic charged current interactions,
\bel{eq:cclep}
{\cal L}\; =\; {g\over \sqrt{2}}\; W_\mu^\dagger\;\sum_{ij} \; 
\bar\nu_{iL}\,\gamma^\mu\, \mathbf{U}_{ij}\, l_{jL} \; +\; \mathrm{h.c.}
\; ,
\ee
involve a flavour mixing matrix $\mathbf{U}$. 
Neglecting possible CP-violating phases, the present data
on neutrino oscillations implies the mixing structure:
$$
\mathbf{U} \; \approx \; \left[  \begin{array}{ccc}
{1\over\sqrt{2}}\, (1+\lambda) & {1\over\sqrt{2}}\, (1-\lambda) & \epsilon \\
-{1\over 2}\, (1-\lambda+\epsilon) & {1\over 2}\, (1+\lambda-\epsilon) &
{1\over\sqrt{2}} \\
{1\over 2}\, (1-\lambda-\epsilon) & -{1\over 2}\, (1+\lambda+\epsilon) &
{1\over\sqrt{2}} \end{array}\right] ,
$$
with $\lambda\sim 0.2$ and $\epsilon < 0.25$ \cite{concha}.
Therefore, the mixing among leptons appears to be very different
from the one in the quark sector.

An important question to be addressed in the future concerns the
possibility of leptonic CP violation and
its relevance for explaining the baryon asymmetry of our universe
through a leptogenesis mechanism \cite{Buchmuller}.

The smallness of the neutrino masses implies a strong suppression
of neutrinoless lepton-flavour-violation processes. This suppression
can be avoided in models with other sources of lepton flavour violation,
not related to $m_{\nu_i}$ \cite{Ma}. The present experimental limits
on lepton-flavour-violating $\tau$ decays,
at the $10^{-6}$ level \cite{CLEOLV,Inami}, are already sensitive to
new-physics scales of the order of a few TeV \cite{BHHS:02}.
Further improvements at future experiments \cite{Santinelli}
would allow to explore interesting and totally unknown phenomena.

\section{SUMMARY}

Our knowledge of the lepton properties has been considerably
improved during the last few years.
Lepton universality has been tested to rather good accuracy,
both in the charged and neutral current sectors. The
Lorentz structure of the leptonic $l\to\nu_l l'\bar\nu_{l'}$ decays
has been determined with good precision in the $\mu$ decay \cite{PDG}
and relevant constraints have been obtained for the $\tau$.
An upper limit of 3.2\% (90\% CL) has been already set on the probability
of having a (wrong) decay from a right-handed $\tau$ 
\cite{taurev98,taurev00}.

The quality of the hadronic $\tau$ decay data
has made possible to perform quantitative QCD tests
and determine the strong coupling constant very accurately,
providing a nice experimental verification of asymptotic freedom.
An improved value of the strange quark mass has been also obtained
from Cabibbo-suppressed hadronic $\tau$ decays.

More recently,
the precise determination of the $\mu$ anomalous magnetic
moment has nearly reached the needed sensitivity to explore
higher-order electroweak corrections.
A further experimental improvement is expected soon.
In order to perform a meaningful precision test of the electroweak theory, 
it is very important to achieve a better control of the QCD contributions.

The Standard Model provides a beautiful theoretical framework which 
is able to accommodate all our present knowledge on electroweak and strong
interactions.
In spite of this impressive phenomenological success, it leaves
too many unanswered questions to be considered as a complete 
description of the fundamental forces. We do not understand yet 
why fermions are replicated in three (and only three)
nearly identical copies. Why the pattern of masses and mixings
is what it is?  Are the masses the only difference among the three
families? What is the origin of the flavour structure?
Which dynamics is responsible for the observed CP violation?

The fermionic flavour is the main source of arbitrary free 
parameters in the Standard Model.
The problem of fermion-mass
generation is deeply related with the mechanism responsible for the 
electroweak spontaneous symmetry breaking.
Thus, the origin of these parameters lies in the most obscure part of
the Standard Model lagrangian: the scalar sector.
Clearly, the dynamics of flavour appears to be ``terra incognita''
which deserves a careful investigation.

The first hints of new physics beyond the Standard Model
have emerged recently, with convincing evidence of 
neutrino oscillations from both solar and atmospheric experiments.
The existence of lepton flavour violation opens a very interesting
window to unknown phenomena. New experiments
will probe the Standard Model to a much deeper level of sensitivity 
and will explore the frontier of its possible extensions.

\section*{ACKNOWLEDGEMENTS}
I would like to thank the organizers for hosting an enjoyable conference.
I want to acknowledge useful discussions with V.~Cirigliano, 
A.~Czarnecki, M.~Davier,
M.C.~Gonz\'alez, F.~Matorras, J.~Portol\'es and P.~Ruiz.
This work has been supported in part by MCYT, 
Spain (Grant FPA-2001-3031), by EU funds for regional development
and by the EU, RTN {\it EURIDICE} (HPRN-CT-2002-00311).


\end{document}